\documentclass[12pt]{article}
\usepackage{amsmath}
\usepackage{amssymb}
\begin{document}
\vskip 2cm
\begin{center}
{\bf {\Large   {Some Novel Features\\ in 2D Non-Abelian Theory: BRST Approach}}}

\vskip 3.0cm

{\sf N. Srinivas$^{(a)}$, S. Kumar$^{(a)}$, B. K. Kureel$^{(a)}$ and  R. P. Malik$^{(a,b)}$}\\
$^{(a)}$ {\it Physics Department, Centre of Advanced Studies,}\\
{\it Banaras Hindu University, Varanasi - 221 005, (U.P.), India}\\

\vskip 0.1cm


$^{(b)}$ {\it DST Centre for Interdisciplinary Mathematical Sciences,}\\
{\it Institute of Science, Banaras Hindu University, Varanasi - 221 005, India}\\
{\small {\sf {e-mails: seenunamani@gmail.com, sunil.bhu93@gmail.com, bkishore.bhu@gmail.com, rpmalik1995@gmail.com}}}
\end{center}

\vskip 2cm
\noindent
{\bf Abstract:}
Within the framework of Becchi-Rouet-Stora-Tyutin (BRST) formalism, we discuss some {\it novel}
features of a two (1+1)-dimensional (2D) non-Abelian 1-form gauge theory (without any interaction with matter
fields). Besides the {\it usual} off-shell nilpotent
and absolutely anticommutating (anti-)BRST symmetry transformations, we discuss the off-shell nilpotent
and absolutely anticommutating (anti-)co-BRST symmetry transformations. Particularly, we lay
emphasis on the existence of the coupled (but equivalent) Lagrangian densities of the 2D non-Abelian theory
in view of the presence of  (anti-)co-BRST symmetry transformations where we pin-point some {\it novel} features
associated with the Curci-Ferrari (CF) type restrictions. We demonstrate that these CF-type
restrictions can be incorporated into the (anti-)co-BRST invariant Lagrangian densities through the {\it fermionic}
Lagrange multipliers which carry specific ghost numbers. The modified versions of the Lagrangian densities
(where we get rid of the {\it new} CF-type restrictions) respect
some precise symmetries as well as a couple of symmetries with CF-type constraints. These observations are
completely novel as far as the BRST formalism, with proper (anti-)co-BRST symmetries, is concerned.

\vskip 2cm
\noindent
PACS numbers: 11.30.Pb, 03.65.-w, 11.30.-j.

\vskip 0.25cm
\noindent
{\it {Keywords}}: {2D non-Abelian theory; nilpotent (anti-)BRST symmetries; nilpotent (anti-) co-BRST symmetries; Curci-Ferrari type restrictions; fermionic Lagrange multipliers}


\section{Introduction}

In modern language, the gauge
theories are characterized by the first-class constraints (in the terminology of Dirac's prescription
for classification scheme of constraints [1, 2]) and their gauge
symmetries are generated by {\it these} constraints. For the covariant canonical quantization of theories, based on the
above gauge symmetries, one of the conceptually elegant, geometrically rich and theoretically beautiful
methods is the Becchi-Rouet-Stora-Tyutin (BRST) formalism where a {\it classical} local
gauge symmetry is traded with a couple of {\it quantum} gauge symmetries which are christened as the BRST
and anti-BRST symmetries. The latter symmetries (i.e. BRST and anti-BRST)
are, however, {\it fermionic} in nature as they are found to be nilpotent of order two. Hence, they are supersymmetric type, too.

We have established, in our recent publications, that a $p$-form ($p= 1, 2, 3,...$) Abelian gauge theory
in $ D = 2p $ dimensions of spacetime respects, in addition to the (anti-)BRST symmetry transformations,
the nilpotent and absolutely anticommuting 
(anti-)co-BRST symmetries, too (see, e.g. [3] and references therein). 
We have been able to demonstrate the existence of the latter type
of symmetries in the 2D (non)-Abelian 1-form gauge theories, too, which has enabled us to establish
that the 2D (non-)Abelian gauge theories (without any interaction with matter fields)
provide a field theoretic example of Hodge theory [4,5] as well as a {\it new} model [6]
of the topological field theory (TFT) which captures a few aspects of Witten-type
of TFTs and some salient features of Schwarz-type TFTs. 

In the above context, it is pertinent to point out that the above {\it basic} (anti-) BRST and 
(anti-)co-BRST symmetries are {\it physically} important because they provide physical realizations [6] of the cohomological
operators of differential geometry and they establish the fact that the Lagrangian densities of 2D non-Abelian theory
(cf. Eqn. (2) below) look like Witten-type TFT but the innate symmetry transformations of this 
theory resemble to that of the symmetry
transformations of Schwarz-type TFTs (see, e.g. [3, 6]). To be more specific, there is no {\it shift} symmetry in our theory which is 
one of the theoretical hallmark of a Witten-type TFT [7]. Rather, the symmetries of our present theory are {\it internal}
in nature [8]. The topological invariants and their recursion relations of our present 2D theory 
have been obtained in our earlier work [6].

As far as discussion on TFTs of the 2D non-Abelain theory is concerned, it has been demonstrated (see. e.g. [6, 3])
that the Lagrangian density as well as symmetric energy-momentum tensor of the theory can be 
expressed as the sum of the anticommutators with BRST and co-BRST charges. This observation is one of the key features
of the Witten type TFT [7]. However, the symmetries of this theory do {\it not} include the shift
symmetry as they are {\it only} internal in nature. Thus, the symmetries of this theory 
capture one of the decisive features of the Schwarz-type TFT [8]. 
These observations establish the novelty in the topological nature of our present theory. Thus, we note
that the existence of the (anti-)BRST and (anti-)co-BRST symmetries is physically 
important and interesting.

The purpose of our present investigation is to point out a few {\it novel} features that
are associated with the (anti-)co-BRST symmetry transformations in the context of  
2D non-Abelian gauge theory (without any interaction with matter fields). We have focused on the coupled 
(but equivalent) Lagrangian densities [9, 10] in the Curci-Ferrari gauge [11, 12] which respect
the proper (anti-)BRST symmetries on the constrained hypersurface (in the 2D spacetime manifold) where the
CF-condition [13] is satisfied (cf. Eqn. (4) below). We have shown that the proof of the {\it equivalence} of the above
coupled Lagrangian densities requires {\it another} set of CF-type restrictions w.r.t. the
(anti-)co-BRST symmetry transformations. We have incorporated these latter CF-type restrictions
in the modified versions of the Lagrangian densities (cf. Eqn. (8) below) which respect the
(anti-)co-BRST symmetry transformations separately and independently. 
We have commented on the origin of these new CF-type restrictions in view
of the existing nilpotent (anti-)BRST and (anti-)co-BRST symmetry transformations of our present 2D non-Abelian theory.

Against the backdrop of the above statements, the central result of our present investigation is the proof of
{\it equivalence} of the coupled Lagrangian densities (2) w.r.t. the (anti-)co-BRST symmetry transformations 
which requires {\it new} type of CF-type restrictions $({\cal B} \times C) = 0$ and  $({\cal B} \times \bar C) = 0$.
The non-trivial solutions of these CF-type constraints are (i) the directions of the ${\cal B}$ and
$C$ fields are parallel in the SU(N) Lie algebraic space for the anti-co-BRST invariance, and (ii) for the 
requirement of the co-BRST invariance, we observe that the direction of the field ${\cal B}$ in the SU(N)
Lie algebraic space must be parallel to the direction of field $\bar C$. These are {\it not} very strong restrictions
as co-BRST and anti-co-BRST symmetries are linearly independent of each-other because of their absolute
anticommutativity property. In otherwords, the CF-type restrictions ${\cal B}\times C = 0$ and ${\cal B}\times \bar C = 0$
do not imply that $C$ and $\bar C$ fields are {\it parallel} to each-other in the Lie-algebraic space.

We call the restrictions $ ({\cal B}\times C) = 0$ and $({\cal B}\times \bar C) = 0$ as CF-type
restrictions from the point of view of symmetries (cf. Eqns. (4) and (7)). However, these restrictions can {\it not} 
be treated on 
equal footing to the original CF- condition $ B + \bar B + (C \times \bar C) = 0$ 
because the latter emerge due to the CF-gauge-fixing in the theory.
To get rid of these new restrictions 
(i.e. $ ({\cal B}\times C = 0, {\cal B}\times \bar C = 0)$, we have incorporated them into the Lagrangian densities (8)
through the Lagrange multiplier fields in such a way that {\it both} the Lagrangian densities (8) respect 
the (anti-)co-BRST symmetry transformations  separately and independently (cf. Eqns. (10) and (11) below).
This is a novel observation in the sense that we can {\it not} perform such kind of exercise with the
{\it usual} CF-condition $ B + \bar B + (C \times \bar C) = 0$ which is associated with the (anti-)BRST 
transformations of our present 2D non-Abelian theory. 

The following key motivating factors have been at the heart of our present investigation. 
First, it is very important for us to establish the {\it equivalence} of the coupled Lagrangian densities
(cf. Eqn. (2) below) w.r.t. the (anti-)co-BRST symmetry transformations because these
symmetries play a very important role in the proof of the 2D non-Abelian theory to be a tractable
field theoretic example of the Hodge theory as well as a {\it new} model of TFT. Second, 
the existence of the CF-type restrictions is one of the key signatures of a {\it quantum} gauge theory
when the latter is discussed within the framework of BRST formalism. We have accomplished
this goal in our present endeavor for the (anti-)BRST as well as (anti-)co-BRST
symmetries. Finally, we have found the modified versions of the coupled
Lagrangian densities (cf. Eqn. (8) below) which respect the (anti-)co-BRST symmetries
separately and independently. 
All the above cited results are {\it novel} as far as the
{\it basic} concepts behind the existence of (anti-)BRST and (anti-)co-BRST 
symmetry transformations for the Lagrangian densities of our 2D non-Abelian theory 
are concerned.

The contents of our present paper are organized as follows. In Sec. 2, we briefly recapitulate the 
bare essentials of the (anti-)BRST symmetry transformations for the 2D non-Abelian
theory so that the convention and notations could be fixed. Our Sec. 3 is devoted to the discussion
of (anti-)co-BRST symmetry transformations and existence of the CF-type restrictions. We comment on {\it all}
the existing continuous symmetries of our theory in their operator form and obtain their algebra in Sec. 4. Finally, 
we make some concluding remarks in Sec. 5.

\section{Preliminaries: (anti-)BRST symmetries}

We begin with the following 2D (anti-)BRST invariant coupled (but equivalent) Lagrangian densities [9, 10]
in the Curci-Ferrari gauge (see, e.g. [11,12] for details) 
\begin{eqnarray}
{\cal L}_B^{(0)}& = &-\frac{1}{4} \,F_{\mu\nu}\cdot F^{\mu\nu} + B\cdot (\partial_{\mu}A^{\mu}) \\ \nonumber
&+& \frac{1}{2}(B\cdot B + \bar B \cdot \bar B)
- i\,\partial_{\mu}\bar C \cdot D^{\mu}C, \\ \nonumber
{\cal L}_{\bar B}^{(0)}& = &-\frac{1}{4} \,F_{\mu\nu}\cdot F^{\mu\nu} - \bar B\cdot (\partial_{\mu}A^{\mu}) \\ \nonumber
&+ &\frac{1}{2}(B\cdot B + \bar B \cdot \bar B) 
- i\, D_{\mu}\bar C \cdot \partial^{\mu}C,
\end{eqnarray}
where the field strength tensor $ F_{\mu\nu} = \partial_{\mu}A_{\nu} - \partial_{\nu}A_{\mu} + i\, (A_\mu\times A_\nu) $
has been defined through the 2-form $F^{(2)} = dA^{(1)} + i \,A^{(1)}\times A^{(1)}$
(with $d = dx^\mu \partial_\mu $ and $d^2 = 0$). The 1-form 
$A^{(1)} = dx^{\mu}\,(A_{\mu}\cdot T)$ defines the vector potential $A_{\mu}^a$ for the 2D non-Abelian 
theory\footnote {We have taken into account the background 2D Minkowski spacetime manifold 
that is endowed with a {\it flat} metric with signatures (+1, -1) and the
Levi-civita tensor $ \varepsilon_{\mu\nu}$ is chosen to be $\varepsilon_{01} = +1 = \varepsilon^{10}$ and 
$\varepsilon_{\mu\nu}\varepsilon^{\nu\lambda} = \delta_{\mu}^\lambda $, etc.,
where the Greek indices $\mu, \nu, \lambda,...=0, 1$. In the SU(N) Lie algebraic space, we define 
the {\it dot} and {\it cross} products between two non-null vectors $P^a$ and $Q^a$ 
 as $P\cdot Q = P^a Q^a$ and $(P\times Q) = f^{abc}\,P^b\, Q^c  T^a$ where 
$a, b, c,... = 1, 2, 3,... N^{2}-1$. The generators $T^a$ satisfy the SU(N) Lie algebra
$[T^a, T^b] = f^{abc}\,T^c$ where the structure constants $f^{abc}$ are chosen to be totally antisymmetric for the semi-simple
Lie group SU(N). 
Throughout the whole body of our text, we denote the nilpotent (anti-) BRST and (anti-)co-BRST symmetry transformations by
the notations $s_{(a)b}$ and $s_{(a)d}$, respectively.}
and the Nakanishi-Lautrup type auxiliary fields $B$ and $\bar B$ obey the CF-condition
$B + \bar B + (C\times \bar C) = 0$ where the fermionic ($ C^a C^b +  C^b C^a = 0, \,
\bar C^a \bar C^b +  \bar C^b \bar C^a = 0, \,\,
C^a \bar C^b + \bar C^b C^a = 0,\, (C^a)^2 = (\bar C^a)^2 = 0 $, etc.) (anti-)ghost fields $(\bar C^a) C^a$ are
required for the validity of unitarity in our theory. The covariant derivative 
$D_{\mu} C = \partial_{\mu} C + i\, (A_{\mu}\times C)$ is in the adjoint representation of the SU(N) Lie algebra of our 
2D non-Abelian theory.\\

For the 2D theory, it is clear that the kinetic term $\bigl[-\,\frac{1}{4} \,(F_{\mu\nu}\cdot F^{\mu\nu})\bigr] $ is equal
to: $ - \frac {1}{2} F_{01}^a \,F^{01a} \equiv \frac {1}{2} E^a E^a $ where
$E^a = F_{01}^a = \partial_{0}A_{1}^a - \partial_{1}A_{0}^{a} + i\, (A_{0}\times A_{1})^a $. This kinetic
term can be linearized through the additional auxiliary field $ \cal B$. The ensuing coupled 
Lagrangian densities are [6,14]:
\begin{eqnarray}
{\cal L}_B &= &{\cal B} {\cdot E} - \frac {1}{2}\,{\cal B} \cdot {\cal B} +\, B\cdot (\partial_{\mu}A^{\mu}) 
\\ \nonumber
&+& \frac{1}{2}(B\cdot B + \bar B \cdot \bar B) 
- i\,\partial_{\mu}\bar C \cdot D^{\mu}C, \\ \nonumber
{\cal L}_{\bar B}& = &{\cal B} {\cdot E} - \frac {1}{2}\,{\cal B} \cdot {\cal B} - \bar B\cdot (\partial_{\mu}A^{\mu}) 
\\ \nonumber
&+& \frac{1}{2}(B\cdot B + \bar B \cdot \bar B) 
- i\, D_{\mu}\bar C \cdot \partial^{\mu}C.
\end{eqnarray}
The above Lagrangian densities respect the following 
(anti-)BRST symmetry transformations $(s_{(a)b})$, namely;
\begin{eqnarray}
&&s_{ab}A_{\mu} = D_{\mu}\bar C,\; s_{ab}\bar C = -\frac{i}{2}\,(\bar C\times \bar C),\; 
s_{ab}C = i\bar B,\\ \nonumber
&&s_{ab}B = i\,(B\times \bar C), \;\;
s_{ab}E = i\,(E\times \bar C),\;\;s_{ab}\bar B = 0,\\ \nonumber
&& s_{ab}{\cal B} = i\,({\cal B}\times \bar C),\;
s_{ab}(\partial_\mu A^{\mu}) = \partial_\mu D^{\mu} \bar C,\;s_b \bar C = i B\\ \nonumber
&&s_b A_{\mu} = D_{\mu}C,\; s_b C = -\frac {i}{2}(C\times C),\;\;
s_b \bar B = i \,(\bar B \times C), \\ \nonumber
&& s_b E = i\,(E\times C), \;\;\; s_b{\cal B} = i\,({\cal B}\times C), \;\;\; s_b B = 0, 
\end{eqnarray}
because the Lagrangian densities ${\cal L}_B $ and ${\cal L}_{\bar B }$ transform under 
$s_{(a)b}$ as (see, e.g. [14] for details):
\begin{eqnarray}
s_b{\cal L}_B &= &\partial_{\mu}[B\cdot D^{\mu}C],\;\;\;
s_{ab}{\cal L}_{\bar B} = \partial_{\mu}[-\bar B\cdot D^\mu \bar C], \\ \nonumber
s_{ab}{\cal L}_ B & = & \partial_{\mu}\bigl[-\{\bar B + (C\times \bar C)\}\cdot \partial^\mu \bar C\bigr ] \\\nonumber
&+ &\bigl[B + \bar B + (C\times \bar C)\bigr ]\cdot D_\mu (\partial^{\mu}\bar C), \\ \nonumber
s_b{\cal L}_ {\bar B}& = &\partial_{\mu}\bigl[\{B + (C\times \bar C)\}\cdot \partial^\mu  C\bigr ]\\\nonumber
&-& \bigl[B + \bar B + (C\times \bar C)\bigr ]\cdot D_\mu (\partial^{\mu} C).
\end{eqnarray}
Thus, we  note that {\it both} the Lagrangian densities are {\it equivalent} in the sense that {\it both}
of them respect the (anti-)BRST symmetry
transformations on a hypersurface in the 2D spacetime manifold where the CF-condition
$B + \bar B + (C\times \bar C) = 0$ is satisfied. In other words, we have 
$s_b {\cal L}_{\bar B} = -\,\partial_\mu\,[\bar B \cdot \partial^{\mu} C]$ and
$s_{ab}{\cal L}_{B} = \partial_\mu\,[B \cdot \partial^{\mu} \bar C]$. \\

We end this section with the following remarks. First, the (anti-)BRST symmetry transformations
$s_{(a)b}$ are off-shell nilpotent $(s_{(a)b}^2 = 0)$ of order two which shows their fermionic
(supersymmetric) nature. Second, the absolute anticommutating property $(s_b s_{ab} + s_{ab} s_b = 0)$
is satisfied by the (anti-)BRST symmetry transformations $s_{(a)b}$ provided the CF-condition
$B + \bar B + (C\times \bar C) = 0$ is invoked [13]. In other words, $s_b$ and $s_{ab}$ have their
own identities on the hypersurface that is defined by the  constraint equation 
$B + \bar B + (C\times \bar C) = 0$ and, therefore, they are
linearly independent of each-other. Third, the coupled Lagrangian densities ${\cal L}_B$
and ${\cal L}_{\bar B}$ are {\it equivalent} only on the above hypersurface as far as the
off-shell nilpotent and absolutely anticommuting (anti-)BRST symmetry transformations $s_{(a)b}$ are concerned.
Finally, we observe that the total kinetic term 
[i.e.$-\frac {1}{4}(F_{\mu\nu}\cdot F^{\mu\nu})\equiv{\cal B}\cdot E - \frac{1}{2} {\cal B}\cdot {\cal B}$],
owing its origin to the exterior derivative
(i.e. $d = dx^\mu \partial_\mu $ and $d^2 = 0$), 
remains invariant under the (anti-) BRST symmetry transformations
of our present 2D theory. To be more specific, we have 
$F^{(2)} = \frac {1}{2} \, (dx^{\mu}\wedge dx^{\nu})\,(F_{\mu\nu}\cdot T) = d A^{(1)} + i\,A^{(1)}\wedge A^{(1)}$
where $A^{(1)} = dx^{\mu}\,(A_\mu \cdot T)$ and $F_{\mu\nu} = \partial_\mu A_\nu - \partial_\nu A_\mu + i\,(A_\mu \times A_\nu)$.
We observe that field strength tensor $F_{\mu\nu}$ owes its origin to the geometrical object
$d = dx^\mu\,\partial_\mu$.\\

\section{Nilpotent (anti-)co-BRST symmetry transformations: Lagrangian formulation}

It is very interesting to note that the Lagrangian densities $ {\cal L}_B $ and $ {\cal L}_{\bar B} $ 
{\it also} respect another set of off-shell nilpotent $(s_{(a)d}^2 = 0)$ and absolutely anticommuting
 $(s_d s_{ad} + s_{ad} s_d = 0)$ (anti-)co-BRST symmetry transformations $s_{(a)d}$, namely;
\begin{eqnarray}
&&s_{ad}A_{\mu} = -\varepsilon_{\mu\nu}\partial^\nu C,\quad s_{ad} C = 0, \quad
s_{ad}\bar C = i\,{\cal B},\\ \nonumber
&&s_{ad}{\cal B} = 0,\qquad s_{ad} E = D_\mu \partial^\mu C,\qquad 
s_{ad}B = 0,\\ \nonumber
&&s_{ad} \bar B = 0, \quad s_{ad}(\partial_\mu A^\mu ) = 0, \qquad
s_d A_{\mu} = -\varepsilon_{\mu\nu}\partial^\nu \bar C,\\ \nonumber
&&s_d \bar C = 0, \qquad s_d C = -i\,{\cal B},\qquad s_d{\cal B} = 0,\qquad s_d B = 0,\\ \nonumber
&& s_d E = D_\mu \partial^\mu \bar C,\qquad 
s_d \bar B = 0, \qquad s_d (\partial_\mu A^\mu ) = 0,
\end{eqnarray}
because we observe that the Lagrangian densities transform to the following {\it total} spacetime derivatives, namely;
\begin{eqnarray}
&&s_d{\cal L}_B = \partial_{\mu}[{\cal B}\cdot \partial^{\mu}\bar C], \quad \qquad
s_{ad}{\cal L}_{\bar B} = \partial_{\mu}[{\cal B}\cdot \partial^\mu  C],
\end{eqnarray}
which imply that the action integrals $( S = \int d^2 x {\cal L}_B $ and $ S = \int d^2 x {\cal L}_{\bar B} )$
remain invariant under $s_d $ and $s_{ad} $ provided these transformations act, separately and independently, on 
${\cal L}_B $ and ${\cal L}_{\bar B}$, respectively. In the above proof of symmetry
invariance, we have used $ \varepsilon_{\mu\nu} (\partial^{\mu}\bar C \times \partial^{\nu}\bar C) = 0 $
as well as $ \varepsilon_{\mu\nu} (\partial^{\mu} C \times \partial^{\nu} C) = 0 $
which are also useful in the proof of the absolute anticommuting and nilpotency properties of $ s_{(a)d} $.

To establish the {\it equivalence}\footnote{By equivalence, we mean exactly similar kind of transformations as we have
obtained in equation (4) for the (anti-)BRST symmetries which demonstrates that ${\cal L}_B $ and ${\cal L}_{\bar B}$
respect {\it both} these symmetries on the hypersurface where $B + \bar B + (C \times \bar C) = 0$.}
of the Lagrangian densities ${\cal L}_B $ and ${\cal L}_{\bar B}$ w.r.t. (anti-)co-BRST symmetry transformations, we observe
that the following transformations are true:
\begin{eqnarray}
s_d{\cal L}_{\bar B}& =& \partial_{\mu}\bigl[{\cal B}\cdot D^{\mu}\bar C  
-\varepsilon^{\mu\nu}(\partial_\nu \bar C\times \bar C)\cdot C\bigr ] \\ \nonumber
&+ &i\,\partial_{\mu}A^\mu\cdot ({\cal B}\times \bar C), \\ \nonumber 
s_{ad}{\cal L}_B &=& \partial_{\mu}\bigl[{\cal B}\cdot D^{\mu} C  
+\varepsilon^{\mu\nu}\bar C\cdot(\partial_\nu  C\times  C)\bigr ] \\ \nonumber 
&+& i\,\partial_{\mu}A^\mu\cdot ({\cal B}\times  C). 
\end{eqnarray}
Thus, we note that {\it both} the Lagrangian densities  ${\cal L}_B $ and ${\cal L}_{\bar B}$
can be {\it equivalent} w.r.t. the (anti-)co-BRST symmetry transformations, iff, we invoke 
the CF-type of restrictions as ${\cal B}\times \bar C = 0,\,{\cal B}\times  C = 0$ 
which are physically {\it allowed} because
both the above restrictions are {\it perfectly} (anti-)co-BRST invariant. In other words, we note that 
$ s_{(a)d}\,[{\cal B}\times \bar C] = 0,\, s_{(a)d}\,[{\cal B}\times  C ]= 0 $. We call 
these restrictions as the CF-type restrictions by taking the analogy from Eqn. (4) where
we have the CF-condition $(B + \bar B + C \times \bar C = 0)$ 
in the context of nilpotent (anti-)BRST symmetry transformations. The above nomenclature has been adopted due to symmetry
considerations {\it only}.

As far as the {\it perfect} (anti-)co-BRST symmetry transformations are concerned, the CF-type restrictions 
$({\cal B}\times \bar C = 0, \, {\cal B}\times  C = 0)$ can be incorporated into the modified
versions of the Lagrangian densities ${\cal L}_B $ and ${\cal L}_{\bar B}$ as  
\begin{eqnarray}
&& {\cal L}_B \rightarrow {\cal L}_{(B,\, \bar\lambda)}  = {\cal B} {\cdot E} 
- \frac {1}{2}\,{\cal B} \cdot {\cal B} +\, B\cdot (\partial_{\mu}A^{\mu}) \nonumber\\
&& +   \frac{1}{2}(B\cdot B + \bar B \cdot \bar B) 
- i\,\partial_{\mu}\bar C \cdot D^{\mu}C 
- \bar\lambda \cdot ({\cal B}\times  C), \nonumber\\
&& {\cal L}_{\bar B} \rightarrow  {\cal L}_{(\bar B, \,\lambda)}  =   {\cal B} {\cdot E} 
- \frac {1}{2}\,{\cal B} \cdot {\cal B} - \bar B\cdot (\partial_{\mu}A^{\mu}) \nonumber\\
&&+   \frac{1}{2}(B\cdot B + \bar B \cdot \bar B) 
- i\, D_{\mu}\bar C \cdot \partial^{\mu}C 
-  \lambda \cdot ({\cal B}\times \bar C), 
\end{eqnarray}
where $ \lambda $ and $\bar\lambda $ are the fermionic (i.e. $ \lambda^2 = 0,\, \bar\lambda^2 = 0, \quad
 \lambda \bar\lambda + \bar\lambda \lambda = 0 $) Lagrange multiplier fields which carry the ghost number
(+1) and (-1), respectively.

It is interesting to note that the above modified version of the Lagrangian density
${\cal L}_{(\bar B,\, \lambda)}$ respects the following {\it perfect} 
(anti-)co-BRST transformations $(s_{(a)d})$:
\begin{eqnarray}
&&s_{ad}A_{\mu} = -\varepsilon_{\mu\nu}\partial^\nu C,\; s_{ad} C = 0,\; 
s_{ad}\bar C = i{\cal B}, \nonumber \\
&&s_{ad}({\cal B}\times \bar C) = 0,
\; s_{ad} E = D_\mu \partial^\mu C,\; s_{ad}\lambda = 0, \\ \nonumber
&&s_{ad}\bigl[{\cal B}, B , \bar B, (\partial_\mu A^\mu )\bigr ] = 0,\;s_d C = -i{\cal B},  \\ \nonumber
&&s_d A_{\mu} = -\varepsilon_{\mu\nu}\partial^\nu \bar C, \;
\; s_d({\cal B}\times \bar C) = 0,\;\; 
s_d E = D_\mu \partial^\mu \bar C,\\ \nonumber
&&s_{d}\lambda = i\,(\partial_\mu A^\mu), \;\;s_d \bar C = 0, \;\;
s_{d}[{\cal B}, B ,  \bar B,  (\partial_\mu A^\mu )] = 0.
\end{eqnarray}
To corroborate on the above statement, we note that the Lagrangian density ${\cal L}_{(\bar B,\,\lambda)}$
transforms as follows:
\begin{eqnarray}
&&s_{ad}{\cal L}_{(\bar B,\, \lambda)} = \partial_{\mu}\bigl[{\cal B}\cdot \partial^{\mu} C \bigr ], \\ \nonumber
&&s_d{\cal L}_{(\bar B, \,\lambda)} = \partial_{\mu}\bigl[{\cal B}\cdot D^{\mu}\bar C  
-\varepsilon^{\mu\nu}(\partial_\nu \bar C\times \bar C)\cdot C \bigr ].
\end{eqnarray}
The above observations establish that the action integral ($S = \int d^2x {\cal L}_{(\bar B ,\, \lambda)}$)
remains invariant under $s_{(a)d}$. In exactly similar fashion, we observe that the Lagrangian density
${\cal L}_{(B,\, \bar \lambda)}$ transforms as
\begin{eqnarray}
&&s_{d}{\cal L}_{(B,\, \bar \lambda)} = \partial_{\mu}\bigl[{\cal B}\cdot \partial^{\mu} \bar C \bigr ],\\ \nonumber
&&s_{ad}{\cal L}_{( B,\, \bar \lambda)} = \partial_{\mu}\bigl[{\cal B}\cdot D^{\mu} C  
+ \varepsilon^{\mu\nu}\bar C\cdot (\partial_\nu  C\times  C\bigr ],
\end{eqnarray}
under the following fermionic (anti-)co-BRST symmetry transformations ($s_{(a)d}$)
\begin{eqnarray}
&&s_{ad} A_{\mu} = -\varepsilon_{\mu\nu}\partial^\nu C,\quad s_{ad} C = 0, \quad
s_{ad}\bar C = i\,{\cal B},\\ \nonumber
&&s_{ad}({\cal B}\times C) = 0,\,
s_{ad} E = D_\mu \partial^\mu C,\, \quad s_{ad} \bar\lambda = i\,(\partial_\mu A^\mu), \\ \nonumber 
&&s_{ad}[{\cal B}, B , \bar B, (\partial_\mu A^\mu )] = 0, \;
s_d A_{\mu} = -\varepsilon_{\mu\nu}\partial^\nu \bar C, \;\; 
s_d \bar C = 0,\\ \nonumber
&&s_d C = -i{\cal B},\;
s_d({\cal B}\times C) = 0,\;\;\;
s_d E = D_\mu \partial^\mu \bar C,\\ \nonumber
&&s_{d}\bar\lambda = 0, \;\;\;
s_{d}[{\cal B}, B ,  \bar B,  (\partial_\mu A^\mu )] = 0,
\end{eqnarray}
which are nilpotent of order two (i.e. $s_{(a)d}^2 = 0$) and absolutely anticommuting
($s_d s_{ad} + s_{ad} s_d = 0 $) in nature. 
As a side remark, we would like to emphasize that it is the gauge-fixing
term (owing its origin to the co-exterior derivative of differential geometry 
$ \delta A^{(1)} = -\, *\, d \,* A^{(1)} = \partial_\mu A^\mu$) that remains invariant
under $s_{(a)d}$. To be more specific, the operator $ \delta  = -\, *\, d \,* $ is the co-exterior derivative where $*$ is the
Hodge duality operation on the 2D flat Minkowski manifold.

At this juncture, we make the following remarks. First, we observe that the CF-type restrictions 
(${\cal B}\times C = 0,\, {\cal B}\times \bar C = 0$), in the context of (anti-)co-BRST symmetry transformations,
are {\it perfectly} invariant under $s_{(a)d}$. Second, the CF-condition  $(B + \bar B + C \times \bar C = 0)$, in 
the context of (anti)BRST transformations ($s_{(a)b}$), transforms under the (anti-)BRST symmetry transformations as: 
\begin{eqnarray}
&&s_b[B + \bar B + (C \times \bar C) ]= i(B + \bar B + C \times \bar C )\times C, \\\nonumber
&&s_{ab}[B + \bar B + (C \times \bar C) ]= i\,(B + \bar B + C \times \bar C )\times \bar C.
\end{eqnarray}
The above observations demonstrate that the 
celebrated CF-condition is (anti-) BRST invariant (cf. Eqn. (13)) {\it only} on the
constrained hypersurface in the 2D spacetime manifold where the constrained field equation
$(B + \bar B + C \times \bar C = 0)$ is valid. Finally, the CF-type restrictions 
(${\cal B}\times C = 0,\, {\cal B}\times \bar C = 0$) could be incorporated
in the Lagrangian densities through Lagrange multipliers $\lambda$ and $\bar \lambda$ 
(cf. Eqn. (8)) but the CF-condition $(B + \bar B + C \times \bar C = 0)$ can {\it not} 
be incorporated in the Lagrangian densities
so that one could have {\it perfect} symmetry
invariance in the theory. In addition to the (anti-)co-BRST symmetry transformations (9), the 
Lagrangian density $ {\cal L}_{(\bar B,\,\lambda)}$ also respects the following anti-BRST symmetry
transformations:
\begin{eqnarray}
&&s_{ab}A_{\mu} = D_{\mu}\bar C,\quad s_{ab}\bar C = -\frac{i}{2}\,(\bar C\times \bar C), \\ \nonumber
&&s_{ab}C = i\bar B,\quad s_{ab}B = i\,(B\times \bar C),\;\;s_{ab}{\cal B} = i\,({\cal B}\times \bar C), \\ \nonumber
&&s_{ab}\lambda = 0, \qquad s_{ab}E = i\,(E\times \bar C),\qquad s_{ab}\bar B = 0. 
\end{eqnarray}
In exactly similar fashion, the Lagrangian density $ {\cal L}_{(B, \,\bar \lambda)}$ respects,
besides the (anti-)co-BRST symmetries (12), the
following BRST symmetry transformations:
\begin{eqnarray}
&&s_b A_{\mu} = D_{\mu}C,\; s_b C = -\frac {i}{2}\,(C\times C),\;s_b \bar C = i B,\\ \nonumber 
&&s_b\bar \lambda = 0,\;\;\;
s_b \bar B = i \,(\bar B \times C), \;\;
s_b E = i\,(E\times C),\; \\ \nonumber
&&s_b{\cal B} = i\,({\cal B}\times C),\;\; s_b B = 0.
\end{eqnarray}
In fact, we observe that $s_{ab} {\cal L}_{(\bar B,\,\lambda)} = \partial_\mu\bigl[-\bar B\cdot D^\mu \bar C\bigr ]$
and $s_b {\cal L}_{( B,\,\bar \lambda)} = \partial_\mu\bigl[B\cdot D^\mu C\bigr ]$.
As a consequence, we note that  $ {\cal L}_{(\bar B, \,\lambda)}$
and $ {\cal L}_{(B, \,\bar \lambda)}$ respect {\it three} perfect\footnote{We do {\it not} invoke any CF-type restriction
for the proof of the symmetry invariance in the theory Thus, the symmetries are {\it perfect}
in the real sense of the word.} continuous 
fermionic symmetries at this stage.

It can be seen that, under the following nilpotent BRST symmetry transformations
\begin{eqnarray}
&&s_b A_{\mu} = D_{\mu}C,\; s_b C = -\frac {i}{2}\,(C\times C),\; s_b \bar C = i B , \\ \nonumber
&& s_b \bar B = i \,(\bar B \times C), \;
s_b E = i(E\times C), \; s_b{\cal B} = i\,({\cal B}\times C), \\ \nonumber
&&s_b B = 0, \quad s_{b}(\partial_\mu A^{\mu}) = \partial_\mu D^{\mu} C,\;
s_b \lambda = -i\,(\lambda \times C),
\end{eqnarray}
the Lagrangian density $ {\cal L}_{(\bar B, \,\lambda)}$ transforms as
\begin{eqnarray}
s_b{\cal L}_ {(\bar B,\,\lambda)} & = &\partial_{\mu}\bigl[\{B + (C\times \bar C)\}\cdot \partial^\mu  C\bigr ]\\ \nonumber
&-& \bigl[B + \bar B + (C\times \bar C)\bigr ]\cdot D_\mu (\partial^{\mu} C)\\ \nonumber
& - & i\,({\cal B}\times \lambda)\cdot \bigl[B + C\times \bar C \bigr ],
\end{eqnarray}
which demonstrate that if we use the CF-conditions $ B + \bar B + C \times \bar C = 0 $ as well as 
${\cal B}\times \bar B = 0$, we obtain the BRST invariance as 
$ s_b{\cal L}_ {(\bar B,\,\lambda)}  = \partial_\mu \bigl [-\bar B\cdot \partial^\mu  C\bigr ]$
because the {\it second} term in (17) is zero due to the CF-condition $(B + \bar B + C \times \bar C = 0) $
and the {\it third} term can be expressed as 
$i\,({\cal B}\times \lambda)\cdot \bar B \equiv -i\,\lambda \cdot ({\cal B}\times \bar B)$ which amounts to 
a CF-type restriction ${\cal B}\times \bar B = 0$. This restriction, however, is also equivalent to
the restrictions $(\bar B \times \lambda) = 0$ and  $({\cal B} \times \lambda) = 0$.

We also observe that the Lagrangian density
$ {\cal L}_{( B, \,\bar\lambda)}$ respects the following nilpotent $(s_{ab}^2 = 0)$ anti-BRST symmetry 
transformations $s_{(ab)}$
\begin{eqnarray}
&&s_{ab}A_{\mu} = D_{\mu}\bar C,\quad s_{ab}\bar C = -\frac{i}{2}\,(\bar C\times \bar C), \\ \nonumber
&&s_{ab}C = i\bar B, \quad
s_{ab}\bar B = 0,\quad s_{ab}B = i\,(B\times \bar C), \\ \nonumber
&&s_{ab}E = i\,(E\times \bar C),\quad 
s_{ab}{\cal B} = i\,({\cal B}\times \bar C),\\ \nonumber
&&s_{ab}(\partial_\mu A^{\mu}) = \partial_\mu D^{\mu} \bar C,\;\;
s_{ab}\bar\lambda = -i\,(\bar \lambda \times \bar C),
\end{eqnarray}
because the Lagrangian density $ {\cal L}_{( B, \,\bar\lambda)}$ transforms as
\begin{eqnarray}
s_{ab}{\cal L}_ {( B,\,\bar\lambda)} & = &\partial_{\mu}\bigl[-\{B 
+ (C\times \bar C)\}\cdot \partial^\mu \bar C\bigr ]\\ \nonumber
&+ &\bigl[B + \bar B + (C\times \bar C)\bigr ]\cdot D_\mu (\partial^{\mu} \bar C)\\ \nonumber
& - & i\,({\cal B}\times \bar\lambda)\cdot \bigl[\bar B + C\times \bar C \bigr ],
\end{eqnarray}
which can become a total spacetime derivative
(i.e. $ s_{ab}{\cal L}_ {( B,\,\bar\lambda)}  = \partial_{\mu}\bigl[B \cdot \partial^\mu \bar C\bigr ] $)
provided we invoke the CF-condition $(B + \bar B + C \times \bar C = 0)$ and $({\cal B}\times B) = 0$. The latter
restriction is also equivalent to the CF-type restrictions $({\cal B}\times \bar \lambda) = 0$ and $(B\times \bar \lambda) = 0$. 
Thus, we require, at least,  {\it two} CF-type restrictions if we wish to have the anti-BRST invariance of
${\cal L}_ {( B,\,\bar\lambda)}$.

We end this section with the remarks on the origin of the CF-type restrictions 
$ {\cal B}\times C = 0,\,{\cal B}\times \bar C = 0,\, {\cal B}\times B = 0$ and ${\cal B}\times \bar B = 0$
which are to be invoked if we wish that the Lagrangian densities ${\cal L}_ {( B,\,\bar\lambda)}$ and
${\cal L}_ {(\bar B,\,\lambda)}$ {\it must} respect {\it all} the {\it four} basic fermionic 
[i.e. (anti-) BRST and (anti-)co-BRST] symmetries of the present 2D non-Abelian theory. It is clear that the
(anti-)BRST invariance of CF-condition $(B + \bar B + C \times \bar C = 0)$ generates {\it no} further
CF-type restriction (cf. Eqn. (13)). However, the requirement of the invariance of {\it this} CF-condition
w.r.t. the (anti-)co-BRST symmetry transformations $s_{(a)d}$ leads to the new  CF-type
restrictions ${\cal B}\times \bar C = 0$ and ${\cal B}\times  C = 0$. These restrictions are {\it perfectly}
(anti-)co-BRST invariant [i.e. $s_{(a)d}({\cal B}\times C) = 0, \,s_{(a)d}({\cal B}\times \bar C) = 0 $]. 
As a consequence, they have been incorporated in the Lagrangian densities through the Lagrange multipliers
so that one can get rid of them. However, these restrictions are {\it not}
(anti-)BRST invariant. The requirements of the (anti-)BRST invariance of these CF-type restrictions lead
to the imposition of new CF-type restrictions ${\cal B}\times \bar B = 0$ and ${\cal B}\times B = 0$
which appear when we demand the (anti-)BRST invariance 
(i.e. $ s_{ab}{\cal L}_ {( B,\,\bar\lambda)}  = \partial_{\mu}\bigl[B \cdot \partial^\mu \bar C\bigr ],  \,
 s_b{\cal L}_ {(\bar B,\,\lambda)}  = \partial_\mu \bigl [-\bar B\cdot \partial^\mu  C\bigr ]$)
of the Lagrangian densities ${\cal L}_ {( B,\,\bar\lambda)}$ and ${\cal L}_ {( \bar B,\,\lambda)}$,
respectively. We note that we do {\it not} need further CF-type of restrictions in our theory as far as 
the symmetry invariances are concerned. 
However, theoretically, we observe that $s_{(a)d}\,[{\cal B}\times B] = 0, \,s_{(a)d}\,[{\cal B}\times \bar B] = 0 $
but the (anti-)BRST invariance of the restrictions $({\cal B}\times B)$ and $({\cal B}\times \bar B)$
leads to new CF-type restrictions $ B\times C = 0$ and $ \bar B\times \bar C = 0$. At this stage, the 
series terminates because the requirements of (anti-)BRST and (anti-)co-BRST invariances produce
{\it no} further new type of CF-type restrictions. The above statements can be mathematically 
expressed as:
\begin{eqnarray}
&&s_d\,\bigl[B+ \bar B + (C\times \bar C)\bigr ] = 0  \; \Longrightarrow   \; {\cal B}\times \bar C = 0, \\ \nonumber
&&s_{ad}\,\bigl[B+ \bar B + (C\times \bar C)\bigr ] = 0 \;  \Longrightarrow   \;  {\cal B}\times C = 0, \\ \nonumber
&&s_b\,\bigl ({\cal B}\times \bar C\bigr ) = 0 \;  \Longrightarrow  \;  {\cal B}\times \bar B = 0, \\ \nonumber
&&s_{ab}\,\bigl ({\cal B}\times  C\bigr ) = 0 \; \Longrightarrow  \;  {\cal B}\times  B = 0, \\ \nonumber
&&s_b\,\bigl ({\cal B}\times  B\bigr ) = 0 \;  \Longrightarrow  \;   B \times  C = 0, \\ \nonumber
&&s_{ab}\,\bigl ({\cal B}\times  \bar B\bigr ) = 0 \;  \Longrightarrow  \; \bar B\times \bar C = 0.
\end{eqnarray}
In the derivation of the above, we have used the CF-condition $(B + \bar B + C \times \bar C = 0)$
as well as other CF-type restrictions that precede the derivation of any new CF-type of restriction
by using the (anti-)BRST symmetry transformations. It should be noted that 
$ s_{(a)d}\,\bigl[{\cal B}\times C, {\cal B}\times \bar C, {\cal B}\times \bar B, {\cal B}\times B \bigr ] = 0 $.
We point out that the requirements of
$ s_{(a)d}\,\bigl[{\cal B}\times C \bigr ] = 0$ and $ s_{(a)d}\,\bigl[{\cal B}\times \bar C \bigr ] = 0 $
lead to the validity of CF-type restrictions ${\cal B}\times \bar B = 0,\, {\cal B}\times  B = 0 $.
We also note and re-emphasize that {\it all} the theoretically allowed CF-type restrictions have {\it not} been
used in our present discussion where we have considered {\it only} the specific kinds of 
infinitesimal and continuous symmetries.

\section{Symmetry operators and their algebra}

In addition to the nilpotent and absolutely anticommuting
(anti-)BRST and (anti-)co-BRST symmetry transformations (18), (16), (15), (14), (12)
and (9), the Lagrangian densities (8) respect the following ghost-scale symmetry transformations 
\begin{eqnarray}
&&C \to e^{\Omega }C,\qquad \bar C \to e^{-\,\Omega }\bar C,\qquad \lambda \to e^{\Omega }\lambda, \\ \nonumber
&&\bar \lambda \to e^{-\,\Omega }\bar\lambda,\qquad \Phi \to e^{0}\,\Phi 
\qquad (\Phi = A_\mu, B, \bar B, {\cal B}, E),
\end{eqnarray}
where $\Omega$ is a global (i.e. spacetime-independent) scale parameter and 
numerals in the exponents denote the ghost numbers for the
corresponding fields. The infinitesimal version (with $\Omega = 1$) of (21) are:
\begin{eqnarray}
&&s_g C = + C,\quad s_g \bar C = -\bar C,\quad s_g \lambda = + \lambda,\\ \nonumber
&&s_g \bar\lambda = - \bar\lambda,\qquad
s_g \Phi =  0,\quad (\Phi = A_\mu, B, \bar B, {\cal B}, E).
\end{eqnarray}
There is a {\it unique} bosonic symmetry in our theory which is obtained from the anticommutators
of the appropriate combinations of $s_{(a)b}$ and $s_{(a)d}$. We define this 
transformation as\footnote{For the sake of brevity,
we have taken here {\it only}  $ s_w = \{s_b, s_d\}$ and have {\it not} 
considered its other definition $ s_w = -\{s_{ab}, s_{ad}\}$.
The latter definition is equivalent to it when we use the EOMs.}
$ s_w = \{s_b, s_d\} \equiv -\{s_{ab}, s_{ad}\}$. The relevant fields of the theory transform, under
$ s_w $, as
\begin{eqnarray}
&&s_w\,A_\mu = \bigl [D_\mu {\cal B} \,+ \, \varepsilon_{\mu\nu}\,\partial^\nu\, B 
+\varepsilon_{\mu\nu}\,(\partial^\nu\bar C\times C) \bigr ], \\ \nonumber
&&s_w \bar B = i \, (\bar B \times {\cal B}), \;\;
s_w\,(\partial_\mu A^\mu)  = \bigl [\partial_\mu D^\mu {\cal B} \\ \nonumber
&&+ \,\varepsilon^{\mu\nu}\,(\partial_\mu C\times \partial_\nu \bar C)\bigr ], \; 
s_w \bigl [\bar C, C, B, {\cal B}, \bar \lambda \bigr ] = 0,\\ \nonumber
&&s_w\,E = \bigl [ i \,(E \times {\cal B}) \,- \,D_\mu\partial^\mu\, B  \\ \nonumber
&&-\,(D_\mu  C \times\,\partial^\mu \bar C) \,- \,(D_\mu \partial^\mu \bar C \times C) \bigr ],
\end{eqnarray}
where $ s_w = \{ s_d, s_b \}$ has been computed from the transformations (12) and (15) of the Lagrangian density 
${\cal L}_{(B, \,\bar \lambda)}$.
Furthermore, the above transformations have been quoted modulo a factor of $ - i$. 
It is straightforward to check that\footnote{There is a simpler way
to compute the expression for $s_w {\cal L}_{(B, \,\bar \lambda)}$. This is due to the fact that $ s_w = \{ s_d, s_b \}$ and the
observations that $s_b {\cal L}_{( B,\,\bar \lambda)} = \partial_\mu\bigl[B\cdot D^\mu C\bigr ]$ 
and $s_{d}{\cal L}_{(B,\, \bar \lambda)} = \partial_{\mu}\bigl[{\cal B}\cdot \partial^{\mu} \bar C \bigr ]$. Thus,
we have $s_w\,{\cal L}_{(B, \,\bar \lambda)} = (s_d s_b + s_b s_d)\, {\cal L}_{(B, \,\bar \lambda)}$ which
leads to the derivation of (24) modulo a factor of $- i$.} 
\begin{eqnarray}
s_w\,{\cal L}_{(B, \,\bar \lambda)}& = & \partial_\mu\,\bigl [ B\cdot D^\mu {\cal B} \, - {\cal B}\cdot \partial^\mu B 
\\ \nonumber
& - &\, {\cal B} \cdot (\partial^\mu \bar C\times C) 
 + \,\varepsilon^{\mu\nu}\, B \cdot (\partial_\nu \bar C \times C) \bigr ].
\end{eqnarray}
The above transformation of the Lagrangian density shows that the action integral $ S = \int d^2 x \,{\cal L}_{(B, \,\bar\lambda)}$
remains invariant under $s_w $.

It is very interesting to emphasize and observe that the above symmetry operators
$s_{(a)b},\,s_{(a)d}, s_w$ and  $s_g $ of our present theory (corresponding to  ${\cal L}_{(B, \,\bar\lambda)}$)
obey the following nice looking algebra: 
\begin{eqnarray}
&&\bigl[s_w,\,s_r \bigr ] = 0, \; (r = b, ab, d, ad, g ), \;\;s_{(a)b}^2 = 0,\\ \nonumber 
&& s_{(a)d}^2 = 0, \;\;
\bigl\{s_d,\,s_b \bigr \} = s_w = -\, \bigl\{s_{ab},\,s_{ad}\bigr \}, \\ \nonumber
&&\bigl\{s_{ad},\,s_d \bigr \} = \bigl\{s_{ab},\,s_b \bigr \} = 
\bigl\{s_b,\,s_{ad} \bigr \} = \bigl\{s_d,\,s_{ab} \bigr \} = 0, \\ \nonumber
&&\bigl[s_g,\,s_b \bigr ] = + \,s_b, \quad \,\bigl[s_g,\,s_{ad} \bigr ] = + \, s_{ad}, \\ \nonumber
&&\,\bigl[s_g,\,s_{ab} \bigr ] = -\, s_{ab},\;\;
\bigl[s_g,\,s_d \bigr ] = - \,s_d.
\end{eqnarray}
The above algebraic structure is true {\it only} when we take into account the validity of equations of motion
and CF-type constraints. For instance, it can be clearly checked that $\{ s_b, s_{ad} \} \, \bar \lambda = 0$ if and only
if the equation of motion $\partial_\mu D^\mu C = 0$ is taken into account. Similarly, we note that
$\{ s_b, s_{ab} \} \, \bar \lambda = 0$ only when we invoke the CF-type restriction ${\cal B} \times \bar \lambda = 0$.
A close look at the above algebra demonstrates that this algebra is reminiscent of the Hodge algebra  
that is obeyed by the de Rham cohomological operators\footnote{On a compact manifold without 
a boundary, we define a set of three ($d, \delta, \Delta $) cohomological 
operators which are known as the exterior derivative $d$ (with $d^2 = 0$), co-exterior derivative
$\delta$ (with $\delta^2 = 0 $) and the Laplacian operator $\Delta = (d + \delta)^2 \equiv \{ d, \delta \}$. These operators are popularly known as the de Rham cohomological operators of differential geometry which obey the Hodge algebra (26) where 
$\delta = \pm\,*\,d\,*$ and $*$ is the Hodge duality operation (see. e.g. [15]).}
of the differential geometry (see, e.g. [15]).
They obey the following well-known algebra [15]:  
\begin{eqnarray}
&& d^2 = 0, \qquad  \delta^2 = 0, \qquad \Delta = (d + \delta)^2, \nonumber\\
&& [\Delta, d ] = 0, \qquad  [\Delta, \delta ] = 0, \qquad \Delta = \{ d , \delta \}.
\end{eqnarray}
It is clear, from the above algebra, that the Laplacian operator behaves like a Casimir 
operator (but {\it not} in the Lie algebraic sense).

A precise comparison between the algebraic relations  (25) and (26) demonstrates that
there is a two-to-one mapping between the continuous symmetry operators and the de Rham
cohomological operators of differential geometry. The explicit form of this mapping is as follows :
\begin{eqnarray}
&&(s_b, s_{ad}) \Rightarrow d, \qquad (s_{d}, s_{ab}) \Rightarrow \delta, \\ \nonumber
&&\{s_d, s_b \} = s_w \equiv - \{s_{ad}, s_{ab} \} \Rightarrow \Delta.
\end{eqnarray}
Thus, we conclude that our present 2D non-Abelian theory is a physical example of Hodge theory because the continuous 
transformations of the theory provide the physical realizations of de Rham cohomological
operators (see, e.g. [6]).

\section{Conclusions}

One of the highlights of our present investigation is the proof of the {\it equivalence} 
of coupled Lagrangian densities (2) w.r.t. the (anti-)co-BRST symmetry transformations
which requires {\it new} CF-type restrictions  ${\cal B}\times C = 0$ and $ {\cal B}\times \bar C = 0$.
These restrictions are quite different from the usual CF-condition $B + \bar B + (C \times \bar C) = 0$
that is required for the absolute anticommutativity of the (anti-)BRST symmetry transformations as well
as for the {\it equivalence} of the Lagrangian densities (2) w.r.t. these symmetries. It is a completely
new observation that the (anti-)co-BRST invariant CF-type restrictions can be incorporated into 
the Lagrangian densities (cf. Eqn. (8)) in  such a way that {\it both} the Lagrangian densities of 
(8) respect the (anti-)co-BRST symmetry transformations separately and independently as is evident
from (10) and (11). Thus, to have the {\it perfect} (anti-)co-BRST symmetries, we get rid of
these {\it new} CF-Type restrictions in our present 2D non-Abelian 1-form gauge theory. 

We have concentrated, in our present investigation, on various aspects of (anti-)co-BRST symmetries
of our 2D non-Abelian gauge theory (without any interaction with matter fields). We have found out that
there are analogues of the CF-type condition in our 2D theory w.r.t. the (anti-)co-BRST symmetries, too
(as is the case with the (anti-)BRST symmetries in the description of non-Abelian theory in any {\it arbitrary}   
dimension of spacetime). There are decisive differences, however, between the CF-condition associated with
the (anti-)BRST symmetries and CF-type restrictions connected with the (anti-)co-BRST symmetry transformations.
We point out some of the decisive differences and a few striking similarities between the CF-type restrictions
from the point of view of symmetries in the next paragraphs of our present section. 

First of all, the CF-condition $(B + \bar B + C \times \bar C = 0)$ transforms under the (anti-)BRST 
symmetry transformations to itself (modulo some cross-product) as is evident from Eqn. (13). On the contrary, 
the CF-type restrictions $({\cal B}\times C = 0, \, {\cal B}\times \bar C = 0)$ transform to {\it zero}
under the (anti-)co-BRST symmetry transformations. Second, the CF-condition is somewhat hidden in
the coupled Lagrangian densities in (1) and/or (2) which respect the (anti-)BRST symmetries. This 
condition (i.e. $B + \bar B + C \times \bar C = 0$) can not be explicitly incorporated in the 
Lagrangian densities through the Lagrange multiplier while maintaining the {\it perfect} (anti-)BRST
symmetries. In contrast, as is evident from the Lagrangian densities (8), the CF-type restrictions
$({\cal B}\times C = 0, \, {\cal B}\times \bar C = 0)$ can be incorporated, through the
fermionic Lagrange multipliers, in such a way that the {\it perfect} (anti-)co-BRST symmetry transformations
(9) and (12) are respected by {\it both} the Lagrangian densities (8), separately and independently.
Finally, the CF-condition $(B + \bar B + C \times \bar C = 0)$ is responsible for the absolute 
anticommuting properties of the (anti-)BRST symmetry transformations. On the contrary, the CF-restrictions
$({\cal B}\times C = 0, \, {\cal B}\times \bar C = 0)$ are {\it not} needed for the proof of the absolute
anticommutativity property of the (anti-) co-BRST symmetry transformations. These are completely novel
observations.

There are a few similarities between the CF-condition associated with the (anti-) BRST symmetry transformations 
and the CF-type restrictions that are intimately connected with the (anti-)co-BRST symmetry transformations.
The CF-condition $(B + \bar B + C \times \bar C = 0)$ emerges when we apply $s_{ab}$ on ${\cal L}_B$ and
$s_{b}$ on ${\cal L}_{\bar B}$ (cf. Eqn. (4)). In exactly similar fashion, the CF-type restrictions
$({\cal B}\times C = 0, \, {\cal B}\times \bar C = 0)$ appear when we apply $s_{ad}$ on ${\cal L}_B$ and
$s_{d}$ on ${\cal L}_{\bar B}$ (cf. Eqn. (7)). It is also obvious that the Lagrangian densities 
${\cal L}_B$ and ${\cal L}_{\bar B}$ (cf. Eq. (2)) turn out to be {\it equivalent} w.r.t. the (anti-)BRST
as well as w.r.t. the (anti-)co-BRST symmetries when we impose the CF-condition $(B + \bar B + C \times \bar C = 0)$
and CF-type restrictions $({\cal B}\times C = 0, \, {\cal B}\times \bar C = 0)$ 
{\it together} from {\it outside}. It is very interesting to note that the Lagrangian densities ${\cal L}_{(B, \bar \lambda)} $ 
and ${\cal L}_{(\bar B, \lambda)} $ (cf. Eqn. (8)) respect {\it three} perfect symmetries but, for the (anti-)BRST
invariances of  these Lagrangian densities, we have to invoke the CF-condition ($ B + \bar B + C \times \bar C = 0 $)
and CF-type restrictions ${\cal B} \times B = 0$ and $ {\cal B} \times \bar B = 0$
(cf. Eqns. (17) and (19)), respectively. The latter CF-type restrictions primarily emerge from the (anti-)BRST invarances
of the CF-type restrictions ${\cal B} \times \bar C = 0$ and $ {\cal B} \times C = 0$. 
It is an attractive idea to incorporate {\it all} the other (anti-)co-BRST invariant CF-type restrictions
(e.g.   ${\cal B} \times B = 0$, $ {\cal B} \times \bar B = 0$, $B \times C = 0$, $ \bar B \times \bar C = 0$)
in the {\it general} form of the coupled (but equivalent) Lagrangian densities and look for the presence of the
{\it perfect} (anti-)BRST and (anti-)co-BRST symmetries.

We end this section with the final remark that the absolute anticommutativity ($\{s_b, s_{ab}\} = 0 $ and 
$\{s_d, s_{ad}\} = 0 $) properties of the (anti-)BRST and (anti-)co-BRST symmetry transformations (cf. Eqn. (25))
imply that the CF-type restrictions  ${\cal B} \times C = 0$, $ {\cal B} \times \bar C = 0$ should {\it not}
be considered {\it together} as $s_d$ and $s_{ad}$ are {\it independent} of each-other. Thus, the above CF-type
restrictions do {\it not} imply that $C$ and $\bar C$ fields are parallel (i.e. $C\times \bar C = 0 $) in the 
Lie-algebraic space. As a consequence, the non-Abelian nature of the theory, with CF-condition 
$(B + \bar B + C \times \bar C = 0)$, is still maintained and it does {\it not} reduce to its Abelian counterpart.
We can {\it not} live without CF-condition [13]
because we have proven, in our earlier
works [16,17], that the existence of the CF-type condition is the signature of a {\it quantum} gauge theory 
(described within the framework of BRST formalism) and it is deeply connected with the geometrical objects
called gerbes (see, e.g. [16, 17] for details). 
It is gratifying that all the tower of possible CF-type restrictions for the 2D non-Abelian theory
(cf. Eqn. (20))
have been derived by using the augmented version of superfield approach to BRST formalism in our recent publication [18].\\


\noindent
{\bf Acknowledgments}\\

\noindent
N. Srinivas and S. Kumar are grateful to the financial support received from the BHU-fellowship under which the 
present investigation has been carried out.

\end{document}